\documentstyle[tighten,preprint,aps,epsfig,amssymb]{revtex}
\preprint{ASI-TPA/12/95}
\title{On Integrable Doebner--Goldin Equations}
\author{P.~Nattermann}
\address{Institute for Theoretical Physics A,
         Technical University Clausthal,\\
         38678 Clausthal--Zellerfeld, Germany, E-mail: {\tt
           aspn@pta3.pt.tu-clausthal.de}}
\author{R.~Zhdanov\thanks{on leave
    from the Institute of Mathematics of the Academy of Sciences of
    Ukraine, Tereshchenkivska Str.~3, 252004 Kiev, Ukraine}}
\address{Arnold-Sommerfeld Institute for Mathematical Physics,
  Technical University Clausthal,\\
   38678 Clausthal--Zellerfeld,
  Germany, E-mail: {\tt asrz@pta3.pt.tu-clausthal.de}}
\def\ds{\displaystyle}
\def\rz{{\mathbb R}}

\def\sfrac#1#2{\mbox{$\frac{#1}{#2}$}}
\def\dt{\partial_t}
\def\x{\vec{x}}
\def\grad{\vec{\nabla}}

\def\Ref#1{(\ref{#1})}
\def\Im{\mbox{Im}}
\def\half{\sfrac{1}{2}}

\begin{document}
\maketitle
\pacs{}
\begin{abstract}
  We suggest a method for integrating sub-families of a family of
  nonlinear {\sc Schr\"odinger} equations proposed by {\sc H.-D.~Doebner}
  and {\sc G.A.~Goldin} in the 1+1 dimensional case which have
  exceptional {\sc Lie} symmetries.
  Since the method of integration
  involves non-local transformations of dependent and independent
  variables, general solutions obtained include implicitly determined
  functions. By properly specifying one of the arbitrary
  functions contained in these solutions, we obtain broad classes of
  explicit square integrable solutions. The physical significance and some
  analytical properties of the solutions obtained are briefly discussed.
\end{abstract}
\section{Introduction}
The semi-direct product of the group of diffeomorphisms and the {\sc
  Abel}ian group of smooth functions on $\rz^n$ may be regarded as a
{\em generalized symmetry group\/} on $\rz^n$. From the representation
theory of this group, {\sc H.-D.~Doebner} and {\sc G.A.~Goldin} derived a
family of nonlinear {\sc Schr\"odinger} equations on $\rz^n$
\cite{DoeGol1,DoeGol2,DoeGol3,DoeGol4}, which have
been called the {\sc Doebner--Goldin}(DG)--equation
\cite{DodMiz1} (For recent progress in the study of these equations
see the contributions in \cite{DoDoNa1:proc}.):
\begin{equation}
  i\hbar\dt \psi = \left(-\frac{\hbar^2}{2m}\Delta+V(\x)\right) \psi
    +i\half \hbar D R_2[\psi]\psi
    +\hbar D'\sum_{j=1}^5 c_j R_j[\psi]\psi,
\label{GDGE}
\end{equation}
where $R_j[\psi]$,\ $j=1,\ldots,5$ are real-valued functionals of the density
$\rho:=\psi\bar\psi$ and the current $\vec{J}=\Im (\bar\psi\grad\psi)$,
\begin{equation}
\begin{array}{c}
\ds  R_1[\psi] := \frac{\grad\cdot\vec{J}}{\rho}
                = \Im\frac{\Delta\psi}{\psi},\quad
  R_2[\psi] := \frac{\Delta\rho}{\rho}
             = \frac{\Delta(|\psi|^2)}{|\psi|^2} ,\quad
  R_3[\psi] := \frac{\vec{J}^2}{\rho^2}
            =\left(\Im\frac{\grad\psi}{\psi}\right)^2, \\[.5ex]
\ds  R_4[\psi] := \frac{\vec{J}\cdot\grad\rho}{\rho^2}
            = \Im\left(\frac{\grad\psi}{\psi}\right)^2,\quad
  R_5[\psi] := \frac{(\grad\rho)^2}{\rho^2}
            =\left(\frac{\grad(|\psi|^2)}{|\psi|^2}\right)^2.
\end{array}
\label{Rj}
\end{equation}
Here the real number $D$ (with the physical dimension of a diffusion
constant) labels unitarily inequivalent representations of the
generalized symmetry group involved in the derivation of the nonlinear
equations (\ref{GDGE}). It has been interpreted as a quantum number
describing dissipative quantum systems \cite{DoeGol3,DoeGol4}.
The real number $D'$ (also with the physical dimension of a diffusion
constant) describes the magnitude of the real
non-linearity and the dimension-less constants $c_j\in\rz$ are \lq
model\rq\ parameters.

For the purpose of this paper it is more convenient to use the
parameterization that is obtained by rewriting
equation \Ref{GDGE} in terms of the real functionals $R_j[\psi]$ only,
following the notation of \cite{DoGoNa1,Goldin6,NatSch1,Natter5}:
\begin{equation}
  F(\nu,\mu):\quad i\dt\psi = i\sum_{j=1}^2 \nu_j R_j[\psi]\psi
                         +\sum_{j=1}^5 \mu_j R_j[\psi]\psi
+ \mu_0 V\psi,\quad \nu_1\neq 0.
\label{nse}
\end{equation}
Particular homogeneous equations of this type have also been considered in the
context of quantum mechanics by other authors, e.g.
\cite{Kibble1,GuePus1,Smolin1,Vigier1,Sabati2,Bertol1,Ushver1}.

One of the interesting features of the family of DG--equations
(\ref{GDGE}) is its invariance under a certain group of
transformations \cite{DoGoNa1,Goldin6}
\begin{equation}
  N_{(\Lambda,\gamma)}(\psi) = \psi^{\half(1+\Lambda+i\gamma)}
    {\bar\psi}^{\half(1-\Lambda +i\gamma)} =
    |\psi|\, e^{i(\gamma\ln|\psi|+\Lambda\arg\psi)},
\label{gt}
\end{equation}
i.e.\ if $\psi$ is a solution of $F(\nu,\mu)$, then
$\psi^\prime=N_{(\Lambda,\gamma)}(\psi)$ is a solution of
$F(\nu^\prime, \mu^\prime)$, where the change of parameters under
$N_{(\Lambda,\gamma)}$ is
\begin{equation}
\begin{array}{c}
\ds
\nu_1^\prime = \frac{\nu_1}{\Lambda},\quad
\nu_2^\prime = -\frac{\gamma}{2\Lambda}\nu_1 +\nu_2,\\[.5ex] \ds
\mu_1^{\,\prime} = -\frac{\gamma}{\Lambda}\nu_1 + \mu_1,\quad
\mu_2^{\,\prime} = \frac{\gamma^2}{2\Lambda}\nu_1-\gamma \nu_2
- \frac{\gamma}{2}\mu_1+\Lambda \mu_2\,,\quad \mu_3^{\,\prime}
 = \frac{\mu_3}{\Lambda},\\[.5ex] \ds
\mu_4^{\,\prime}= -\frac{\gamma}{\Lambda}\mu_3 + \mu_4,\quad
\mu_5^{\,\prime} = \frac{\gamma^2}{4\Lambda}\mu_3
- \frac{\gamma}{2}\mu_4
+ \Lambda\mu_5,\quad
\mu_0^{\,\prime} = \Lambda \mu_0.
\end{array}
\label{pt}
\end{equation}
Thus, without loss of generality we can restrict our calculations to
the particular choice of parameters (a particular {\em gauge\/}, see
below)
\begin{equation}
  \nu_1=-1,\qquad \nu_2=0.
\label{gauge}
\end{equation}

Since the transformations \Ref{gt} leave the position probability
invariant, i.e.\ $\rho^\prime(\x,t) = \rho(\x,t)$, they have been called {\em
nonlinear gauge
transformations} \cite{DoGoNa1,Goldin6,DoGoNa2}. This notion is
physically motivated by the fact, that in (non-relativistic) quantum
mechanics we basically measure positions at different
times. Furthermore, the transformations have been used to construct a
consistent notion of observables in a nonlinear quantum theory
\cite{Luecke1}.

It turned out that besides such important properties of the DG--equation
as homogeneity, separability, and Euclidean invariance, which were \lq
input\rq\ by construction, equations (\ref{GDGE}) possess a number of
other attractive properties. Among them one should emphasize the
possibility of constructing explicit square integrable solutions, which
is important for a physical interpretation.
In particular, some stationary and non-stationary ({\sc Gauss}ian and
traveling wave) solutions have been obtained
\cite{DoeGol4,DodMiz1,NatSch1,Goldin3,NaScUs1,NaScUs3}.

The well-known
connection between exact solutions of partial differential equations
(PDEs) and their symmetry properties
\cite{Ovsian1:book,Olver1:book,FuShSe1:book} as well as the
necessity of classifying equations (\ref{GDGE}) in a unified way,
motivated a systematic study of their {\sc Lie} symmetry in
\cite{Natter5}. As a result, one has to distinguish nine sub-families
(characterized by conditions on the parameters $\mu$ in the chosen
gauge) with different maximal {\sc Lie} symmetry algebras $sym_.^.(n)$.
The relationship between these sub-families and their symmetries is
indicated in Fig.1 (using the notation of \cite{Natter5}).

\begin{figure}[tb]
\begin{center}
\epsfig{file=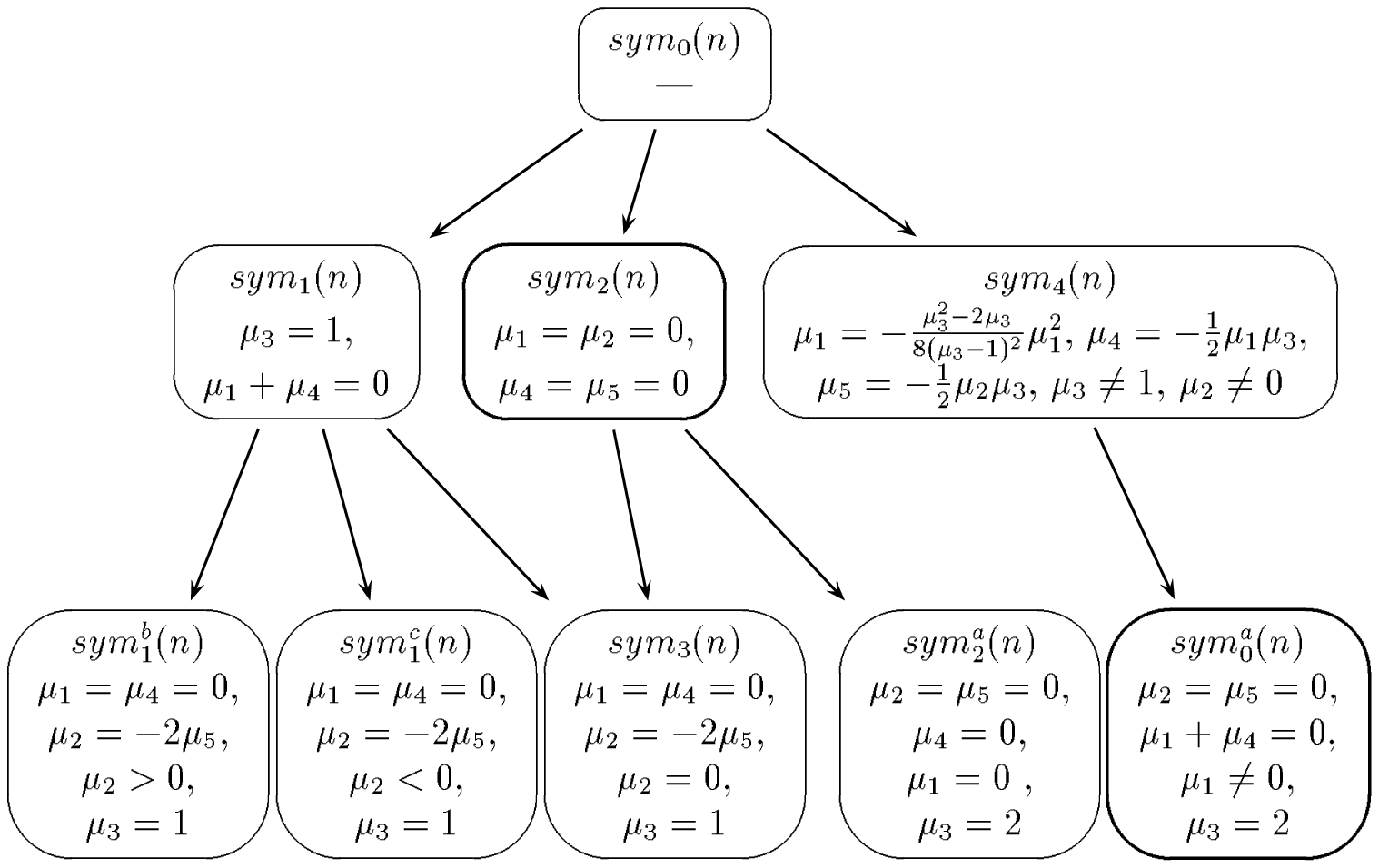}
\end{center}
\begin{quote}
Fig.~1: {\sc Lie} symmetries of the
  DG--equation. Sub-families are characterized by their parameters
  and arrows indicate the subfamily--structure. The equations dealt with in
  this paper are in bold frames.
\end{quote}
\end{figure}

Five of these symmetry algebras are finite dimensional, $sym_0(n)$ --
$sym_4(n)$. Among
them we find the direct sum of the (centrally
extended) {\sc Schr\"odinger} algebra and the real numbers (due to real
homogeneity of the equations). These equations thus fit into the classes of
{\sc
  Schr\"odinger} invariant nonlinear evolution equations determined in
\cite{FusChe1,FusChe2,FusChe3,RidWin1}.

The four remaining symmetry algebras are infinite
dimensional. $sym_1^b(n)$
and $sym_1^c(n)$ contain in addition to the elements of $sym_1(n)$
infinite dimensional algebras $b$ and $c$, that
depend on a pair of (real) solutions of a linear forward and backward heat
equations and a (complex) solution of a linear {\sc Schr\"odinger}
equation, respectively. Actually these symmetries correspond to
linearizations of these sub-families, the first to a pair of
forward and backward heat equations, the latter to a {\sc
  Schr\"odinger} equation \cite{AubSab1,Natter2}.
On the contrary, the symmetry algebras
$sym^a_2(n)$ and $sym^a_0(n)$ contain an infinite dimensional algebra
$a$ that depends only on one real-valued function.
As a consequence, there is no local transformation
(i.e.\ a transformation that does not involve integrals or derivatives of the
dependent variables) linearizing the
corresponding DG--equations. Nevertheless, these equations as well as the
one admitting the finite dimensional symmetry algebra $sym_3(n)$ are shown
in the present paper to be integrable by a {\em non-local\/}
transformation of dependent and independent variables in the case of
one spatial variable ($n=1$).  Thus, {\em all\/}
DG--equations with exceptional symmetries (bottom row of
Fig.~1) $sym_1^b$,\ $sym_1^c$,\
$sym_3$,\ $sym_2^a$,\ $sym_0^a$ are {\em integrable}, i.e.\
they can be reduced to an equation which is either linear or
integrable by quadratures.

The principal object of study in the present paper are
DG--equation in 1+1 dimensions with parameters
\begin{eqnarray}
&&\nu_1=-1,\quad \nu_2=0,\quad \mu_1=\mu_2=\mu_4=\mu_5=0,\label{p1}\\
&&\nu_1=-1,\quad \nu_2=0,\quad\mu_2=\mu_5=0,\quad \mu_3=2,\quad
\mu_4=-\mu_1\ne 0,\label{p2}
\end{eqnarray}
i.e. the following coupled two-dimensional PDEs:
\begin{eqnarray}
  i\psi_t&=&\left\{-i \Im\, {\psi_{xx}\over\psi}
          +\mu_3 \left(\Im {\psi_x\over\psi}\right)^2+
          \mu_0V(x)\right\}\psi,\label{e1}\\[1ex]
  i\psi_t&=&\left\{(\mu_1-i) \Im\, {\psi_{xx}\over\psi}
          +2\left(\Im {\psi_x\over\psi}\right)^2 -
          \mu_1\Im\left({\psi_x\over\psi}\right)^2+
          \mu_0V(x)\right\}\psi.\label{e2}
\end{eqnarray}
One of these DG-equations is contained in the so-called {\sc
  Ehrenfest} sub-family \cite{DoeGol4,Natter2} fulfilling the second
{\sc Ehrenfest} relation: equation \Ref{e1} with $\mu_3=1$, i.e.\ the
DG--equation with maximal {\sc Lie} symmetry $sym_3(n)$. This equation
is furthermore the only {\sc Schr\"odinger} and therefore
{\sc Galilei} invariant equation among \Ref{e1} and
\Ref{e2}. Nevertheless, in the free case ($V\equiv 0$) all of these
DG-equations admit traveling (solitary) wave solutions with arbitrary
shape \cite{NatSch1}. These solutions are rediscovered as a particular
case of the general solutions in this paper.

Using a polar decomposition
\begin{equation}
\label{ansatz}
  \psi(x,t) = \exp\left(r(x,t)+is(x,t)\right)
\end{equation}
we rewrite the above equations in the following way:
\begin{eqnarray}
  \label{f1}
  &F_1:&\left\{
  \begin{array}{rrcl}
    \ds r_t  + s_{xx} +2 r_x s_x &=& 0, \\[.5ex]
    \ds  s_t +\mu_3 s_x^2 &=& -\mu_0 V,
  \end{array}\right.\\[1ex]
\label{f2}
  &F_2:&\left\{
  \begin{array}{rrcl}
    \ds r_t   + s_{xx} +2 r_x s_x &=& 0, \\[.5ex]
    \ds  s_t + \mu_1 s_{xx} + 2 s_x^2 &=& -\mu_0 V.
  \end{array}\right.
\end{eqnarray}
The paper is organized as follows. In the section \ref{2:sec} we integrate the
free equations, i.e. equations (\ref{f1}), (\ref{f2}) with a vanishing
potential ($V\equiv 0$). In order to integrate $F_1$ we have to
distinguish between the cases $\mu_3\neq 1$ and $\mu_3 = 1$, the latter
corresponding to the subfamily with the larger {\sc Lie} symmetry
algebra $sym_3(n)\supset sym_2(n)$.

Section \ref{3:sec} contains some remarks on the integration of the equations
with potential and two particular examples where the integration is
carried out. The methods of integration of $F_1$ and
$F_2$ in sections \ref{2:sec} and \ref{3:sec} yield their general
solutions containing implicitly determined function. Consequently,
these solutions are,
generally speaking, implicit.  Therefore, in section \ref{4:sec} we give some
explicit
solutions for the free equation as well as for linear and quadratic
potentials by specifying one of the arbitrary functions of the general
solutions obtained in the preceding sections.

\section{Integration of free DG--equations}\label{2:sec}
Putting in (\ref{f1}), (\ref{f2}) $V=0$ we obtain the following PDEs:
\begin{eqnarray}
  \label{1.1}
  &\widetilde F_1:&\left\{
  \begin{array}{rrcl}
    \ds r_t  + s_{xx} +2 r_x s_x &=& 0, \\[.5ex]
    \ds  s_t +\mu_3 s_x^2 &=&0,
  \end{array}\right.\\[1ex]
\label{1.2}
  &\widetilde F_2:&\left\{
  \begin{array}{rrcl}
    \ds r_t   + s_{xx} +2 r_x s_x &=& 0, \\[.5ex]
    \ds  s_t + \mu_1 s_{xx} + 2 s_x^2 &=&0.
  \end{array}\right.
\end{eqnarray}
Henceforth we suppose that in (\ref{1.2}) $\mu_1\ne 0$, since
otherwise system (\ref{1.2}) is a particular case of (\ref{1.1}) with
$\mu_3=2$.

\subsection{Integration of the family $\widetilde F_1$}\label{2a:sec}

First, we turn to the integration of the system of nonlinear PDEs
(\ref{1.1}). As this system admits only a finite-dimensional {\sc Lie}
symmetry group, there is no local transformation which linearizes
it. So the only possibility to transform the system in question into an
integrable form is to utilize a non-local transformation of
dependent and independent variables. The choice of a desired change of
variables is implied by the form of the second equation of the
system (\ref{1.1}); it is nothing but the one-dimensional
{\sc Hamilton--Jacobi} equation, which is known to be linearizable by the
following contact transformation (called in the literature the
{\sc Euler-Amp\'ere} transformation):
\begin{equation}
z_0=t,\quad z_1=s_x,\quad u=xs_x-s,\quad
  u_{z_0}=-s_t,\quad u_{z_1}=x.
\label{1.3}
\end{equation}
Let us recall that a transformation $\big(x,t,s(x,t)\big) \mapsto
\big(z_0,z_1,u(z_0,z_1)\big)$,
\begin{eqnarray}
  z_\alpha&=&f_\alpha(x,t,s,s_x,s_t),\nonumber\\
      u&=&g(x,t,s,s_x,s_t),\label{1.4}\\
  u_{z_\alpha}&=&h_\alpha(x,t,s,s_x,s_t),\nonumber
\end{eqnarray}
where $\alpha = 0,1$, is called {\em contact}, if it preserves the
first-order tangency condition
\begin{displaymath}
  ds - s_x dx - s_t dt = 0\quad \Longrightarrow\quad du - u_{z_0}dz_0
  -u_{z_1}dz_1 = 0.
\end{displaymath}
The above condition ensures that the functions $u_{z_0},u_{z_1}$
determined by the last two formulae from (\ref{1.4}) are really
derivatives of a function $u$ determined by the third formula with
respect to $z_0$ and $z_1$ determined by the first and second formula,
correspondingly.

It is readily seen that formulae (\ref{1.3}) determine a contact
transformation preserving the tangency condition. But before applying
the {\sc Euler-Amp\'ere} transformation to the system under study we must
ensure its invertibility, as we may loose some solutions otherwise. It
is known that transformation (\ref{1.3}) is invertible in a domain
where $s_{xx}\ne 0$. Consequently, we have to consider the cases
$s_{xx}\not \equiv 0$ and $s_{xx}\equiv 0$ separately.

\begin{description}
\item[Case 1.] {$s_{xx}\not \equiv 0$}\\

Let us apply the transformation (\ref{1.3}) to
the system (\ref{1.1})
having prolonged it to the second derivatives
\begin{equation}
  \label{1.5}
  u_{z_0z_0}={s_{tx}^2-s_{tt}s_{xx}\over s_{xx}},\quad
  u_{z_0z_1}=-{s_{tx}\over s_{xx}},\quad
  u_{z_1z_1}={1\over s_{xx}}.
\end{equation}
As a result, we get
\begin{equation}
  \label{1.6}
  u_{z_0}=\mu_3 z_1^2,\quad
  r_{z_0}+{2z_1-u_{z_0z_1}\over u_{z_1z_1}}r_{z_1} = - {1\over u_{z_1z_1}}.
\end{equation}
Now, the first equation becomes linear and is easily integrated to
give the following expression for $u(z_0,z_1)$:
  \begin{equation}
    \label{1.7}
      u = \mu_3 z_0z_1^2 + f(z_1),\qquad f\in C^2(\rz,\rz).\\
  \end{equation}
Inserting the result into the second equation of the system
(\ref{1.6}) we get a first-order linear PDE with non-constant coefficients,
\begin{equation}
  \label{1.8}
  r_{z_0} + {2(1 - \mu_3)z_1\over 2\mu_3 z_0 + f''(z_1)}r_{z_1} = -
  {1\over 2\mu_3 z_0 + f''(z_1)}.
\end{equation}
 When integrating the above equation we have to distinguish two sub-cases
$\mu_3 = 1$ and $\mu_3\ne 1$. Let us recall that the DG--equation with
parameters (\ref{p1}) under $\mu_3 = 1$ satisfies the {\sc Ehrenfest}
relation and, what is more, admits an additional symmetry operator
(see Fig.1).

\item[Sub-case 1.1.] {$\mu_3 = 1$}\\

In this case equation (\ref{1.8}) takes the form
\begin{displaymath}
   r_{z_0}= - {1\over 2z_0 + f''(z_1)}
\end{displaymath}
and its general solution reads
\begin{equation}
  \label{1.9}
  r(z_0,z_1) = - \sfrac{1}{2} \ln\Bigl(f''(z_1) + 2z_0\Bigr) + g(z_1),
  \qquad g\in C^2(\rz,\rz) .
\end{equation}
To rewrite the result obtained in the initial variables
$\big(x,t,r(x,t),s(x,t)\big)$ we have to invert the transformation
(\ref{1.3}). From the second and third relations it follows that
\begin{displaymath}
  s = xz_1 - u = xz_1 - tz_1^2 - f(z_1).
\end{displaymath}
To determine the function $z_1=z_1(x,t)$ we make use of the last
relation from (\ref{1.3}). Substituting into it the formula
(\ref{1.7}) with $\mu_3 = 1$ yields
\begin{displaymath}
  x = u_{z_1}=2z_0z_1 + f'(z_1),
\end{displaymath}
hence
\begin{equation}
  \label{1.10}
  2tz_1 - x + f'(z_1)=0.
\end{equation}
The above relation determines the function $z_1(x,t)$ in an implicit
way. Since $s_{xx}=2t+f''(z_1)\not\equiv 0$, we can always solve
(\ref{1.10}) (at least locally) with respect to $z_1$ thus getting an
explicit form of the function $z_1$.

Summarizing the results we conclude that the general solution
of the system of PDEs (\ref{1.1}) with $s_{xx}\not \equiv 0,\ \mu_3 = 1$
is of the form
\begin{equation}
\left\{
\begin{array}{rcl}
  r(x,t) &=& - \sfrac{1}{2} \ln\Bigl(f''(z_1) + 2t\Bigr) +
  g(z_1),\\[1.5ex]
  s(x,t) &=&\ds -tz_1^2 + xz_1 - f(z_1),
\end{array}\right.
\label{1.11}
\end{equation}
where $f,g\in C^2(\rz,\rz)$ are arbitrary functions
and $z_1=z_1(x,t)$ is determined by the relation (\ref{1.10}).

\item[Sub-case 1.2.] {$\mu_3 \ne 1$}\\

Using in (\ref{1.8}) the transformation
\begin{displaymath}
  r(x,t) = \tilde r(x,t) - \frac{1}{2(1-\mu_3)} \ln z_1
\end{displaymath}
we get
\begin{displaymath}
\tilde r_{z_0} - \frac{2(1-\mu_3)z_1}{2\mu_3 z_0 +
  f''(z_1)}\tilde r_{z_1} = 0.
\end{displaymath}
{}From the theory of the first-order PDEs it is well-known (see,
e.g. \cite{CouHil1:book}) that a general solution of the above equation has the
form $\tilde r = g\big(\omega(z_0,z_1)\big)$, where $\omega(z_0,z_1)$ is an
integral of the {\sc Euler-Lagrange} system
\begin{displaymath}
{dz_0\over 1}={dz_1\over {2(1 -\mu_3)z_1\over 2\mu_3 z_0 + f''(z_1)}}.
\end{displaymath}

The above system is rewritten as a linear first-order ordinary
differential equation for a function $z_0=z_0(z_1)$,
\begin{displaymath}
{dz_0\over dz_1} = {1\over 2z_1(1 - \mu_3)}\Bigl(2\mu_3 z_0 +
  f''(z_1)\Bigr)
\end{displaymath}
the general solution of which can be represented in the form
\begin{displaymath}
  C = 2(\mu_3 - 1)z_0z_1^{\frac{\mu_3}{\mu_3-1}} + \int^{z_1}
  \zeta^{\frac{1}{\mu_3 - 1}}f''(\zeta)d\zeta
\end{displaymath}
with an arbitrary constant $C$. Hence we conclude that the general
solution of equation (\ref{1.8}) is given by the following formula:
\begin{equation}
  \label{1.12}
  r(x,t) = \frac{1}{2(\mu_3 - 1)} \ln z_1 + g\left(2(\mu_3 -
    1)z_0z_1^{\frac{\mu_3}{\mu_3-1}} + \int^{z_1} \zeta^{\frac{1}{\mu_3 -
      1}}f''(\zeta)d\zeta\right),
\end{equation}
where $f,g\in C^2(\rz,\rz)$ are arbitrary functions.

Returning to the initial variables $x,t,r,s$ we obtain the
general solution of DG--equation (\ref{1.1}) for the case $s_{xx}\not
\equiv 0,\ \mu_3\ne 1$
\begin{equation}
\left\{
\begin{array}{rcl}
  r(x,t) &=&\ds \frac{1}{2(\mu_3 - 1)}\ln z_1 + g\left(2(\mu_3 -
    1)tz_1^{\frac{\mu_3}{\mu_3-1}} + \int^{z_1} \zeta^{\frac{1}{\mu_3 -
    1}}f''(\zeta)d\zeta\right),\\[1.5ex]
  s(x,t) &=& - \mu_3 t z_1^2 + x z_1 - f(z_1),
\end{array}\right.
\label{1.13}
\end{equation}
where $f,g\in C^2(\rz,\rz)$ are arbitrary functions
and the $z_1=z_1(x,t)$ is determined implicitly
\begin{equation}
  2\mu_3 t z_1 - x + f'(z_1) = 0.
  \label{1.14*}
\end{equation}
\item[Case 2.] {$s_{xx}\equiv 0$}\\

In this case $s(x,t)=\alpha(t)x + \beta(t)$ with arbitrary smooth
functions $\alpha(t)$,\ $\beta(t)$. Substituting this expression into
the second equation from (\ref{1.1}) we arrive at the relations:
\begin{displaymath}
  \alpha'(t)x + \beta'(t) + \mu_3\alpha^2(t)=0,\quad
  r_t+2\alpha(t)r_x=0.
\end{displaymath}
An integration of these equations gives rise to the following
expressions for $r$ and $s$:
\begin{equation}
\left\{
\begin{array}{rcl}
  r(x,t) &=& f(x - 2C_1t),\\[1.5ex]
  s(x,t) &=& C_1x - \mu_3C_1^2t + C_2,
\label{1.14}
\end{array}\right.
\end{equation}
where $C_1,C_2$ are arbitrary real constants. Thus we have
rediscovered the traveling (solitary) wave solutions with arbitrary
shape \cite{NatSch1} as a particular case of the general
solution.
\end{description}
We have established that any smooth solution is contained (at
least locally) in one of the classes given by equations
(\ref{1.11}), (\ref{1.13}), and (\ref{1.14}). Summarizing
we arrive at the conclusion that the general solution of the free
DG--equation (\ref{e1}) splits
into two inequivalent classes:
\begin{enumerate}
\item $\underline{\mu_3 = 1}$
  \begin{eqnarray}
    \psi(x,t)&=&f(x - 2C_1t)\exp\biggl\{i(C_1x - C_1^2t +
    C_2)\biggr\},\label{1.15}\\
    \psi(x,t)&=&\Bigl(f''(z_1) +
    2t\Bigr)^{-\frac{1}{2}}g(z_1)\exp\biggl\{-i\Bigl(tz_1^2 - xz_1 +
    f(z_1)\Bigr)\biggr\},\label{1.16}
  \end{eqnarray}
  where $f,g$ are arbitrary sufficiently smooth functions, $C_1,C_2$ are
  arbitrary real parameters, and $z_1=z_1(x,t)$ is determined implicitly by
  formula (\ref{1.10});
\item $\underline{\mu_3 \ne 1}$
  \begin{eqnarray}
  &&\begin{array}{rcl}
    \psi(x,t)&=&\ds
      f(x - 2C_1t)\exp\biggl\{i(C_1x - \mu_3C_1^2t + C_2)\biggr\},
    \end{array} \label{1.17}\\
  &&\begin{array}{rcl}
    \psi(x,t)&=&\ds
      z_1^{\frac{1}{2(\mu_3-1)}}g\biggl(2(\mu_3 -1)t
      z_1^{\frac{\mu_3}{\mu_3-1}} + \int^{z_1} \zeta^{\frac{1}{\mu_3
          -1}} f''(\zeta)d\zeta\biggr)\\
      &&\times\exp\biggl\{-i\Bigl(\mu_3tz_1^2 - xz_1 +
      f(z_1)\Bigr)\biggr\},
    \end{array} \label{1.18}
  \end{eqnarray}
  where $f,g$ are arbitrary sufficiently smooth functions, $C_1,\ C_2$ are
  arbitrary real parameters, and $z_1=z_1(x,t)$ is determined implicitly by
  formula (\ref{1.14*}).
\end{enumerate}

Although formulae (\ref{1.17}) and (\ref{1.18}) give the general
solution of the corresponding DG--equation for all $\mu_3\ne 1$, the
case $\mu_3=0$ deserves a special consideration, as
the system of PDEs \Ref{1.1} with $\mu_3=0$ is easily integrated without
applying the contact transformation (\ref{1.3}), (\ref{1.5}). The
second equation of (\ref{1.1}) yields that $s$ does not depend on time, so
$$
  s(x,t)=f(x),\qquad f\in C^2(\rz,\rz).
$$
Inserting this into the first equation we get a first order PDE for
$r$,
\begin{equation}
  \label{1.18b}
  r_t + 2f'(x)r_x + f''(x) = 0.
\end{equation}
The case $s(x,t)\equiv const$ leads to a traveling wave solution
(\ref{1.17}) with $\mu_3=0,\ C_1=0$; if
$f'(x)\not\equiv 0$, then the general solution reads
\begin{equation}
  \label{1.18c}
  r(x,t)=g\left(2t - \int^x{d\xi\over f'(\xi)}\right)
  -\sfrac{1}{2}\ln f'(x),
\end{equation}
where $g\in C^2(\rz,\rz)$ is again an arbitrary function.

Consequently, the general solution of the DG--equation (\ref{e1}) with
$\mu_3=0$ is either given by a traveling wave solution (\ref{1.17}) with
$\mu_3=0,\ C_1=0$, or by
\begin{eqnarray}
  \psi(x,t)&=&\Bigl(f'(x)\Bigr)^{-\frac{1}{2}} g\left(2t - \int^x
  {d\xi\over f'(\xi)}\right) \exp\{if(x)\}.
\label{1.18d}
\end{eqnarray}

\subsection{Integration of the family $\widetilde F_2$}\label{2b:sec}

Let us turn to the integration of the DG--equation (\ref{1.2}).
First, we note that the second equation is the potential Burgers
equation, which is is linearized by the logarithmic substitution.
Furthermore, we reduce the order of spatial derivatives in the first
equation by a linear transformation of the dependent variables. Thus,
the transformation
\begin{equation}
\label{1.19}
  v(x,t) = -\mu_1 r(x,t)+s(x,t),\quad u(x,t) =
  \exp\left(\frac{2}{\mu_1} s(x,t)\right),
\end{equation}
reduces system (\ref{1.2}) to the form
\begin{equation}
  \label{1.21}
  uv_t + \mu_1u_x v_x = 0,\quad u_t + \mu_1u_{xx} = 0.
\end{equation}
The second equation may be taken as the integrability condition of the
vector-field $( u,-\mu_1u_x)$ on space--time, ${\partial_t} u = {\partial_x}
(-\mu_1u_x)$, so that it is the gradient of a smooth function $\varphi(x,t)$,
\begin{equation}
  \label{1.22}
  \varphi_t = -\mu_1 u_x,\quad \varphi_x = u.
\end{equation}
With this remark the first equation of the system (\ref{1.21}) is
rewritten to be
\begin{displaymath}
  \varphi_xv_t - \varphi_tv_x = 0
\end{displaymath}
and is easily integrated $v(x,t)=f\big(\varphi(x,t)\big)$, where $f$ is
arbitrary, sufficiently smooth function.

Solving (\ref{1.22}) with respect to $\varphi$ we get
\begin{equation}
  \label{1.23}
  \varphi(x,t) = \int_{0}^{x}u(\xi,t)d\xi -
  \mu_1 \int_{0}^{t}u_x(0,\tau)d\tau + C,
\end{equation}
where $C$ is an arbitrary constant.

Returning to the initial variables $\big(r(x,t),s(x,t)\big)$ we get the
general solution of the DG--equation (\ref{e2})
\begin{equation}
  \label{1.25}
  \psi(x,t) = \big(u(x,t)\big)^{\frac{1}{2}}f\biggl( \int_{0}^{x}
      u(\xi,t)d\xi - \mu_1 \int_{0}^{t}u_x(0,\tau)d\tau\biggr)
      \exp\biggl\{\frac{i\mu_1}{2}\ln u(x,t)\biggr\},
\end{equation}
where $u(x,t)$ is an arbitrary solution of the heat equation
\begin{equation}
  u_t + \mu_1 u_{xx} = 0,
\label{1.24}
\end{equation}
and $f$ is an arbitrary smooth function.

Finally, we note that the traveling wave solutions of
\cite{NatSch1} are reobtained using the particular solution
\begin{displaymath}
  u(x,t) = \exp\left\{ -\frac{v}{\mu_1}\left(x-vt\right)\right\}
\end{displaymath}
of the heat equation \Ref{1.24}.

\section{DG--equation with non-vanishing potential}\label{3:sec}

Surprisingly enough, DG--equations (\ref{f1}), (\ref{f2}) are
integrated in quadratures even in the case when $V(x)\ne 0$ (i.e. in
the presence of a non-vanishing potential). Unfortunately, for the family
$F_1$ the corresponding formulae are implicit and cumbersome. That is
why we restrict ourselves to considering in detail system (\ref{f1}) with an
additional constraint $\mu_3=1$ (i.e.\ the {\sc Ehrenfest} subfamily
of (\ref{f1}) is studied); this system was also considered in
\cite{AubSab1}.

\subsection{Integration of the family $F_1$}\label{3a:sec}
Choosing in (\ref{f1}) $\mu_3=1$, we obtain the following system of
PDEs:
\begin{equation}
  \label{1.26}
   r_t + 2r_x s_x + s_{xx}=0,\quad s_t +  s_x^2 + \mu_0V(x) = 0.
\end{equation}
To linearize this system we make use of the following
trick: instead of system of PDEs (\ref{1.26}) one of its differential
consequences is considered
\begin{equation}
  \label{1.27}
   r_t + 2r_x s_x + s_{xx}=0,\quad s_{tx} +  2s_xs_{xx} + \mu_0V'(x) =
   0.
\end{equation}
Substituting in (\ref{1.27})
\begin{equation}
\label{1.27a}
  R(x,t)=r(x,t),\quad S(x,t)=s_x(x,t)
\end{equation}
we arrive at the system of first-order PDEs
\begin{equation}
  \label{1.28}
   R_t + 2SR_x + S_{x}=0,\quad S_{t} +  2SS_{x} + \mu_0V'(x) =
   0.
\end{equation}
Thus, using the substitution (\ref{1.27a}) enables us to
reduce the order of system of PDEs under study.
Next, we apply to this system the hodograph transformation
$\big(x,t,R(x,t),S(x,t)\big) \mapsto
\big(z_0,z_1,v(z_0,z_1),u(z_0,z_1)\big)$
\begin{equation}
\begin{array}{c}
\ds t = u ,\quad x = z_1,\quad S = z_0,\quad R = v,\\[1pt]
  R_t = \frac{v_{z_0}}{u_{z_0}},\quad
  R_x = v_{z_1} - \frac{u_{z_1}}{u_{z_0}}v_{z_0},\quad
  S_t = {1\over u_{z_0}},\quad
  S_x = -\frac{u_{z_1}}{u_{z_0}}.
\end{array}
\label{1.29}
\end{equation}
This hodograph transformation is defined in an arbitrary
domain where $S_t\not\equiv 0$. So again we have to distinguish two
cases, $S_t \not\equiv 0$ and $S_t\equiv 0$.

\begin{description}
\item[{\bf Case 1.}] $S_t\not\equiv 0$\\

Performing in (\ref{1.28}) the change of variables (\ref{1.29}) we
get
\begin{displaymath}
  \mu_0V'(z_1)u_{z_0} - 2z_0u_{z_1} + 1 = 0,\quad
  \mu_0V'(z_1)v_{z_0} - 2z_0v_{z_1} + {u_{z_1}\over
    u_{z_0}} = 0.
\end{displaymath}
A further change of variables
\begin{equation}
  \label{1.29a}
  \tilde u(z_0,z_1) = u(z_0,z_1),\quad \tilde v(z_0,z_1) = v(z_0,z_1)
  + \sfrac{1}{2}\ln u_{z_0}(z_0,z_1)
\end{equation}
transforms the system to
\begin{equation}
\label{1.30}
  \mu_0V'(z_1)\tilde u_{z_0} - 2z_0\tilde u_{z_1} + 1 = 0,\quad
  \mu_0V'(z_1)\tilde v_{z_0} - 2z_0\tilde v_{z_1}     = 0.
\end{equation}
Thus, combining local and non-local transformations of the dependent
and independent variables we {\em linearized\/} and decoupled the differential
consequence of (\ref{1.26}). Integrating these equations yields
\begin{equation}
\label{1.31}
\begin{array}{rcl}
\tilde u(z_0,z_1)&=&\ds\sfrac{1}{2}\int^{z_1} \left(z_0^2 +\mu_0\Bigl(V(z_1)
-V(\zeta)\Bigr)\right)^{-\frac{1}{2}}d\zeta
    +f\Bigl(z_0^2 + \mu_0 V(z_1)\Bigr),\\[.5ex]
\tilde v(z_0,z_1)&=&g\Bigl(z_0^2 + \mu_0V(z_1)\Bigr),
\end{array}
\end{equation}
where $f,g\in C^2(\rz,\rz)$ are arbitrary functions.

Returning to the variables $\Bigl(x,t,R(x,t),S(x,t)\Bigr)$ we obtain
the general solution of system (\ref{1.28}) in an implicit form
\begin{eqnarray}
  t&=&\sfrac{1}{2} \int^x \left(S^2(x,t) + \mu_0\Bigl(V(x) -
    V(\xi)\Bigr)\right)^{-\frac{1}{2}}d\xi
    + f\Bigl(S^2(x,t) + \mu_0V(x)\Bigr),
\label{1.32a}\\
  R(x,t)&=& \sfrac{1}{2}\ln S_t(x,t) + g\Bigl(S^2(x,t) +
\mu_0V(x)\Bigr).
\label{1.32b}
\end{eqnarray}
To rewrite these equations in the initial dependent variables
$\big(r(x,t),s(x,t)\big)$ we have to invert the transformation
(\ref{1.27a}). As a result, we get
\begin{equation}
\label{1.33}
  r(x,t)=R(x,t),\quad s(x,t)= \int_0^x S(\tau,t)d\tau +
  \varphi(t),
\end{equation}
where $\varphi\in C^2(\rz,\rz)$ is an arbitrary function.
The ambiguity arising is  connected to the fact that we are not
solving the initial system (\ref{1.26}) but its differential
consequence (\ref{1.27}). The \lq extra \rq\ function $\varphi(t)$ is
used to choose from the set of solutions of system (\ref{1.28})
those ones which satisfy (\ref{1.26}).
Indeed, substituting formulae (\ref{1.33}) into (\ref{1.26}) and taking into
account that the functions $R(x,t),S(x,t)$ satisfy system of PDEs
(\ref{1.27}) we arrive at the following ordinary differential equation
for a function $\varphi(t)$:
\begin{displaymath}
  \varphi' + S^2(0,t) + \mu_0V(0) = 0,
\end{displaymath}
so
\begin{displaymath}
 \varphi(t)=-\int_0^tS^2(0,\tau)d\tau -\mu_0V(0)t + C,
\end{displaymath}
where $C$ is an arbitrary real constant.

Summing up, we conclude that the general solution of the initial
DG--equation reads
\begin{equation}
\left\{
\begin{array}{rcl}
r(x,t)&=&\ds -\sfrac{1}{2}\ln S_t(x,t) + g\Bigl(S^2(x,t) +
\mu_0V(x)\Bigr), \\[1.5ex]
s(x,t)&=&\ds \int_0^x S(\tau,t)d\tau - \int_0^t
S^2(0,\tau)d\tau -\mu_0V(0)t + C,
\end{array}
\right.
\label{1.34}
\end{equation}
where $S(x,t)$ is a smooth function determined implicitly by
\Ref{1.32a} and $f,g$ are arbitrary sufficiently smooth functions.

\item[{\bf Case 2.}] $S_t\equiv 0$\\

With this condition the system of PDEs (\ref{1.28}) is easily integrated
to yield
\begin{displaymath}
  \begin{array}{rcl}
  R(x,t)&=&\ds -\sfrac{1}{4}\ln \big(C_1 - \mu_0V(x)\big) + g\Big(2t -
   \ds\int^x{d\xi\over \sqrt{C_1 - \mu_0V(\xi)}}\Big),\\
  S(x,t)&=&\ds\sqrt{C_1 - \mu_0V(x)},
  \end{array}
\end{displaymath}
where $g$ is an arbitrary sufficiently smooth function and $C_1$ is an
arbitrary real constant.

Rewriting the above expressions in the initial variables
$\big(r(x,t),s(x,t)\big)$ we get
\begin{equation}
\left\{
\begin{array}{rcl}
  r(x,t)&=&\ds  -\sfrac{1}{4}\ln \big(C_1 - \mu_0V(x)\big) + g\Big(2t -
   \ds\int^x{d\xi\over \sqrt{C_1 - \mu_0V(\xi)}}\Big),\\[1.5ex]
  s(x,t)&=& \ds \int_0^x \sqrt{C_1-\mu_0(\xi)}d\xi -C_1t+C_2\,,
\end{array}
\right.
\label{1.37}
\end{equation}
where $C_2$ is an arbitrary constant.
\end{description}

Thus, we have established that the general solution of DG--equation
(\ref{e1}) with $\mu_3=1$ splits into the following two classes:
\begin{enumerate}
\item $S_t\not\equiv 0$
\begin{equation}
\begin{array}{rcl}
  \psi(x,t)&=&\ds
  \Bigl(S_t(x,t)\Bigr)^{\frac{1}{2}}g\Bigl(S^2(x,t) +\mu_0V(x)\Bigr)\\
  &&\ds
  \times\exp\left\{i\left(\int_0^x S(\xi,t)d\xi -
  \int_0^t S^2(0,\tau)d\tau -\mu_0V(0)t +
  C\right)\right\},
\end{array}
\label{1.38}
\end{equation}
where $g\in C^2(\rz,\rz)$, $C\in\rz$, and $S(x,t)$ is determined
implicitly by \Ref{1.32a}.
\item $S_t\equiv 0$
\begin{equation}
\begin{array}{rcl}
  \psi(x,t)&=&\ds
  \Bigl(C_1 - \mu_0V(x)\Bigr)^{-\frac{1}{4}} g\left(2t -
  \int_0^x{d\xi\over \sqrt{C_1 - \mu_0V(\xi)}}\right)\\
  &&\ds
  \times\exp\left\{i\left(\int_0^x\sqrt{C_1-\mu_0V(\xi)}d\xi
  - C_1t + C_2\right)\right\},
\end{array}
\label{1.39}
\end{equation}
where $g\in  C^2(\rz,\rz)$ and $C_1,C_2\in\rz$.
\end{enumerate}

\subsection{Integration of the family $F_2$}\label{3b:sec}

Integrating system (\ref{f2}) with potentials is similar to
integrating the free system ($V(x)=0$) in section \ref{2b:sec}.
Using the change of variables (\ref{1.19}) for (\ref{f2}) we arrive at
the following system of PDEs for new functions $u(x,t)$ and $v(x,t)$:
\begin{equation}
  u_t + \mu_1u_{xx} + \frac{2\mu_0}{\mu_1}V(x)u = 0,\quad
  uv_t + \mu_1u_xv_x + \mu_0 V(x)u = 0.
\label{1.40}
\end{equation}
Now, given a potential $V(x)$ and an arbitrary solution $u(x,t)$ of the
first equation of this system, one can construct a general
solution $v(x,t)$ of the second equation which leads to a general
solution of the initial system:
\begin{equation}
  \psi(x,t) = \Bigl(u(x,t)\Bigr)^{\frac{1}{2}}\exp\left(
    -\mu_1^{-1}v(x,t) +i \frac{\mu_1}{2}\ln u(x,t)\right) .
\label{1.41}
\end{equation}

\section{Explicit solutions}\label{4:sec}

As mentioned before some of the solutions of
DG--equation obtained are {\em local\/} in a sense that they are not
determined on the whole plane $\rz^2$. But for physical
applications one needs {\em global\/} solutions, and what is more, they
should be square integrable, i.e. the integral
\begin{equation}
  p=\int_{-\infty}^{\infty}\bar\psi(x,t)\psi(x,t) dx
\label{2.1}
\end{equation}
is to be finite. If it is, the quantity $\rho(x,t)=\frac{1}{p}
\bar\psi(x,t)\psi(x,t) \ge 0$ is treated as a probability density of a
distribution of the wave function $\psi$ in space at a given time.

\subsection{Explicit solutions of the family $F_1$}
Evidently, the traveling wave solutions of DG--equation (\ref{e1}) given
by (\ref{1.15}), (\ref{1.17}) are defined on the whole plane and,
consequently, are global. To ensure square integrability of these
solutions one has to restrict the choice of the arbitrary function $f$
to square integrable ones,
\begin{eqnarray*}
  p&=&\int_{-\infty}^{\infty}f^2(\tau) d\tau <\infty.
\end{eqnarray*}
Thus, the traveling wave solutions are square integrable provided $f$
is.

Solutions (\ref{1.16}), (\ref{1.18}) are, generally speaking,
local, since the function $z_1(x,t)$ contained in these
solutions is determined implicitly by formulae (\ref{1.10}) and
(\ref{1.14*}), correspondingly, and the existence of solution is only
guaranteed locally by the implicit function theorem.
In order to obtain explicit expressions for global and strictly local
solutions we consider solutions of
(\ref{1.16}) with quadratic and cubic functions $f$, respectively.

For quadratic functions $f$,
\begin{displaymath}
  f(z_1) = \mu_3 \alpha z_1^2\,,
\end{displaymath}
the implicit equation \Ref{1.14*} (resp.~\Ref{1.10}) for $z_1$ can be solved
globally and we get $z_1(x,t)=\frac{x}{2\mu_3( t-\alpha)}$.
Thus, we arrive at the following class of explicit solutions of the
DG--equation (\ref{e1})
containing an arbitrary smooth function $g$:
\begin{enumerate}
\item \underline{$\mu_3=1$}
  \begin{equation}
    \label{2.4}
    \psi(x,t)=(2t)^{-\frac{1}{2}}g\left(\frac{x}{2(t-\alpha)}\right)
    \exp\left\{\frac{ix^2}{4(t-\alpha)}\right\},
  \end{equation}
\item \underline{$\mu_3\neq 1$}
  \begin{equation}
    \psi(x,t) = (t-\alpha)^{\frac{1}{2(1-\mu_3)}} x^{\frac{1}{2(\mu_3-1)}}
    g\left((t-\alpha)^{\frac{1}{\mu_3-1}} x^{\frac{\mu_3}{\mu_3-1}}\right)
    \exp\left \{{ix^2\over 2\mu_3 (t-\alpha)}\right\}.
   \label{2.7}
   \end{equation}
\end{enumerate}
These solutions are square integrable, provided $g$ is, and are well
defined on the whole plane $\rz^2$ with a possible exception of the
line $t=\alpha$, where they converge to a $\delta$-function. In
particular for $g(z)=\exp(-z^2)$  solutions \Ref{2.4} coincide with the
{\sc Gauss}ian wave solutions of \cite{NaScUs1}.

Cubic functions $f$,
\begin{displaymath}
  f(z_1) = \sfrac{1}{3} z_1^3\,,
\end{displaymath}
give rise to strictly local solutions of the DG--equation. Indeed,
inserting them into (\ref{1.10}) we obtain a quadratic equation with
respect to $z_1$
\begin{equation}
\label{2.5}
  z_1^2 + 2\mu_3 t z_1 - x =0.
\end{equation}
It has real solutions in the case $x+\mu_3^2t^2\ge 0$ only, i.e.\
$z_1(x,t)$ is not defined inside the parabola $x+\mu_3^2t^2 = 0$.
Solving (\ref{2.5}) yields $z_1(x,t) = -t\pm\sqrt{x+\mu_3^2t^2}$;
according to the general solutions \Ref{1.16} and \Ref{1.18} we have
to choose the positive sign since in \Ref{1.16} $2t+f''(z_1)$ and in
\Ref{1.18} $z_1$ have to be positive.
Hence, we arrive at the following class of strictly local explicit
solutions of the family $F_1$ \Ref{e1} containing an arbitrary smooth
function $g$:
\begin{enumerate}
\item \underline{$\mu_3=1$}
  \begin{equation}
    \psi(x,t) = (x+t^2)^{-\frac{1}{4}} g\left(-t+\sqrt{x+t^2}\right)
    \exp\left\{-i\Bigl(\frac{2}{3}t^3 + tx -\frac{2}{3}
      (x+t^2)^{\frac{3}{2}} \Bigr)\right\}\,,
  \label{2.6}
  \end{equation}
\item \underline{$\mu_3\neq 1$}
  \begin{equation}
  \begin{array}{rcl}
    \psi(x,t)&=&
      \left(\sqrt{x+\mu_3^2t^2}-\mu_3\right)^{\frac{1}{2(\mu_3-1)}}
      g\biggl(\frac{2(\mu_3 -1)}{2\mu_3-1}
      \left(\sqrt{x+\mu_3^2t^2}-\mu_3\right)^{\frac{\mu_3}{\mu_3-1}}\\
      && \times\Bigl((\mu_3-1) t +\sqrt{x+\mu_3^2t^2}\Bigr)\biggr)\\
      &&\ds \times\exp\left\{\frac{i}{3}\left(2(x +\mu_3^2t^2)(-\mu_3
      +\sqrt{x+\mu_3^2t^2}) +\mu_3 t x \right)\right\} \,.
\end{array}
\end{equation}
\end{enumerate}
The domain of definition of these solutions is the set
$\{(x,t): x+\mu_3^2 t^2> 0\}$. Furthermore, as the function
$\sqrt{x+\mu_3^2t^2}$ is not defined for $x$ at $-\infty$ at any
given time $t$, $|t|<\infty$, the solution (\ref{2.6}) is not square
integrable.

In this context let us remark that in general solutions (\ref{1.16})
are square integrable at a given time $t$ provided
\begin{itemize}
\item the (possibly infinite) limits $a_{\pm}=\ds\lim_{x\to
      \pm\infty}z_1(x,t)$ exist and
\item $g$ is square integrable on the interval $[a_-,\, a_+]$.
\end{itemize}
This statement follows from a change of the
integration variable $x\to z_1(x,t)$.

Before turning to DG--equations with non-vanishing potentials we examine
solution \Ref{1.18d} of the particular case $\mu_3=0$ of the family
$F_1$. If the first derivative of $f$ has no zeros, then the
solution given by (\ref{1.18d}) is certainly global. Again, a change
of the integration variable shows that the solution is square
integrable, provided that
\begin{itemize}
\item the (possibly infinite) limits
    $a_{\pm}=\ds\lim_{\pm\infty}\int_0^x{d\tau\over
      f'(\tau)}$ exist and
\item $g$ is square integrable on the interval  $[a_-,a_+]$.
\end{itemize}

In case of non-vanishing potentials we concentrate on the following
specific potentials:
\begin{enumerate}
\item the linear potential
  \begin{equation}
    V(x)=\frac{a}{\mu_0} x,\qquad a\in\rz\,;
  \label{2.8a}
  \end{equation}
\item the harmonic oscillator potential
  \begin{equation}
    V(x)={a^2\over\mu_0}x^2,\qquad a\in\rz\,;
  \label{2.8b}
  \end{equation}
\item the anti-harmonic oscillator potential
  \begin{equation}
    V(x)=-{a^2\over\mu_0}x^2,\qquad a\in\rz\,;
  \label{2.8c}
  \end{equation}
\end{enumerate}
First we consider {\sc Ehrenfest} case $\mu_3=1$, the integration of which
has been studied in detail in section \ref{3a:sec}.
\begin{enumerate}
\item
For linear potentials \Ref{2.8a} the implicit equation (\ref{1.32a})
reads
\begin{equation}
  \label{2.12}
  t = -\frac{1}{a}S(x,t) + f(S^2(x,t) + ax).
\end{equation}
If we choose $f\equiv 0$, then $S(x,t)=-at$. Thus, we get a class of
explicit solutions from \Ref{1.38}:
\begin{eqnarray}
  \psi(x,t)&=&g(x + at^2)\exp\left\{-i\Bigl(atx +\frac{a^2}{3}t^3
  - iC\Bigr)\right\}\label{2.13}.
\end{eqnarray}
These solutions are defined on the whole plane $\rz^2$ and square
integrable, provided $g$ is.

Another class of explicit solutions is obtained directly by means of
formula (\ref{1.39}):
\begin{equation}
\psi(x,t)=(C_1-ax)^{-\frac{1}{4}}g\left(at+\sqrt{C_1-ax}\right)
         \exp\left\{i\Bigl(-\frac{2}{3a}(C_1-ax)^{\frac{3}{2}} - C_1t +
            C_2\Bigr)\right\},
\label{2.14}
\end{equation}
where $g$ is an arbitrary twice continuously differentiable function,
$C_1,\ C_2$ are arbitrary parameters. These solutions are defined on
the half-plane $\{(x,t): x< \frac{C_1}{a} \}$.
\end{enumerate}
Analogously we construct explicit solutions for the (anti-)harmonic
oscillator potentials. We give these without derivation. ($f$ is an
arbitrary sufficiently smooth function, $C_1,\ C_2$ are arbitrary
parameters.)
\begin{enumerate}
\setcounter{enumi}{1}
\item harmonic oscillator potential \Ref{2.8b}
  \begin{eqnarray}
  \psi(x,t) &=& {x}^{-\frac{1}{2}}f\left({x^{-1} \sin
      2at}\right) \exp\left\{i\Bigl(\frac{a}{2}x^2\cot 2at + C_1\Bigr)
      \right\},
  \label{2.15}\\[.5ex]
  \psi(x,t)&=&
  \begin{array}[t]{l}
    \ds(C_1^2-a^2x^2)^{-\frac{1}{4}}f\left
      (\sqrt{C_1^2-a^2x^2}\sin 2at - ax \cos 2at\right)\\
    \ds\times\exp\left\{i\Bigl(\frac{x}{2}\sqrt{C_1^2-a^2x^2}
     +\frac{C_1^2}{2a} \arcsin {ax\over C_1} -C_1t +C_2\Bigr)\right\};
  \end{array}\label{2.16}
  \end{eqnarray}
\item anti-harmonic oscillator potential \Ref{2.8c}:
  \begin{eqnarray}
  \psi(x,t)&=&{x}^{-\frac{1}{2}}f\left({x^{-1}\sinh 2at}\right)
      \exp\left\{i\Bigl(\frac{a}{2}x^2\coth 2at + C_1\Bigr)\right\},
  \label{2.17}\\[.5ex]
  \psi(x,t)&=&\begin{array}[t]{l}
    \ds(C_1+a^2x^2)^{-\frac{1}{4}}f\left(\left(ax
      +\sqrt{C_1+a^2x^2}\right) {\rm e}^{-2at} \right) \\
    \ds \times\exp\left\{i\Bigl(\frac{x}{2}\sqrt{C_1+a^2x^2} +
      \frac{C_1}{2a} \ln \left|ax + \sqrt{C_1+a^2x^2}\;\right| -C_1t +
      C_2\Bigr)\right\}.
  \end{array}\label{2.18}
  \end{eqnarray}
\end{enumerate}

As mentioned in section \ref{3a:sec}, general solutions of the family
$F_1$ (\ref{e1}) with $\mu_3\ne 1$ are given by cumbersome implicit
formulae. But with the particular choice of the potentials above it has
explicit solutions containing one arbitrary function:
\begin{itemize}
\item[I.] $\mu_3=0$
  \begin{enumerate}
  \item linear potential \Ref{2.8a}:
    \begin{equation}
      \label{2.19}
      \psi(x,t)= g(x+at^2)\exp\{-iatx\}\,;
    \end{equation}
  \item harmonic oscillator potential \Ref{2.8b}:
    \begin{equation}
      \label{2.20}
      \psi(x,t)= \mbox{e}^{a^2t^2}g\left(x\mbox{e}^{2a^2t^2}\right)
      \exp\{-ia^2tx^2\}\,;
    \end{equation}
  \item anti-harmonic oscillator potential \Ref{2.8c}:
    \begin{equation}
      \label{2.21}
      \psi(x,t)= \mbox{e}^{-a^2t^2}g\left(x\mbox{e}^{-2a^2t^2}\right)
      \exp\{ia^2tx^2\};
    \end{equation}
  \end{enumerate}
\item[II.] $\mu_3=\lambda^2 > 0$
  \begin{enumerate}
  \item linear potential \Ref{2.8a}:
    \begin{equation}
      \label{2.22}
      \psi(x,t)= g(x+at^2) \exp\left\{-i\Bigl(atx +
        \frac{a^2\lambda^2}{3}t^3 - C\Bigr)\right\};
    \end{equation}
  \item harmonic oscillator potential \Ref{2.8b}:
    \begin{equation}
      \label{2.23}
      \psi(x,t)=x^{-\frac{1}{2}} g\left(x^{-\lambda^2}\sin
        2a\lambda t\right) \exp\left\{i\Bigl(\frac{a}{2\lambda}
      x^2\cot 2a\lambda t +C\Bigr)\right\};
    \end{equation}
  \item anti-harmonic oscillator potential \Ref{2.8c}:
    \begin{equation}
      \label{2.24}
      \psi(x,t)=x^{-\frac{1}{2}}g\left(x^{-\lambda^2}\sinh
          2a\lambda t\right) \exp\left\{i\Bigl(\frac{a}{2\lambda}
        x^2\coth 2a\lambda t +C\Bigr)\right\};
    \end{equation}
  \end{enumerate}
\item[III.] $\mu_3=-\lambda^2 < 0$
  \begin{enumerate}
  \item linear potential \Ref{2.8a}:
    \begin{equation}
      \label{2.25}
      \psi(x,t)=g(x+at^2)\exp\left\{-i\Bigl(atx -
        \frac{a^2\lambda^2}{3}t^3 -C\Bigr)\right\};
    \end{equation}
  \item harmonic oscillator potential \Ref{2.8b}:
    \begin{equation}
      \label{2.26}
      \psi(x,t)=x^{-\frac{1}{2}}g\left(x^{\lambda^2}\cosh
          2a\lambda t\right) \exp\left\{i\Bigl(\frac{a}{2\lambda}
        x^2\tanh 2a\lambda t +C\Bigr)\right\};
    \end{equation}
  \item anti-harmonic oscillator potential \Ref{2.8c}:
    \begin{equation}
      \label{2.27}
      \psi(x,t)=x^{-\frac{1}{2}}g\left( x^{\lambda^2}\cos 2a\lambda
        t\right) \exp\left\{i\Bigl(\frac{a}{2\lambda} x^2\tan
        2a\lambda t +C\Bigr)\right\}.
    \end{equation}
  \end{enumerate}
\end{itemize}
In all these cases $f\in C^2(\rz,\rz)$ and $C\in\rz$.

\subsection{Explicit solutions of the family $F_2$}
Clearly, if the function $u(x,t)$ is a global solution of the heat
equation (\ref{1.24}), then the formula (\ref{1.25}) gives a global
solution of DG--equation (\ref{e2}). And what is more, it is square
integrable provided
\begin{itemize}
\item the (possibly infinite) limits $a_{\pm}
  =\ds\lim_{\pm\infty}\left(\int_0^xu(\tau,t)
    d\tau - \mu_1\int_0^tu_x(0,\tau) d\tau\right)$ exist and
\item $f$ is square integrable on the interval $[a_-,a_+]$.
\end{itemize}

In order to construct explicit solutions of the family $F_2$ \Ref{f2}
for linear and quadratic potentials (\ref{2.8a})--(\ref{2.8c}) we have
to solve equations \Ref{1.40}. After some tedious calculations we obtain the
following solutions:
\begin{enumerate}
  \item linear potential \Ref{2.8a}:
    \begin{equation}
      \label{2.29}
      \psi(x,t)=f(x+at^2)\exp\left\{-iatx - \frac{2ia}{3}t^3
        +iC\right\}\,;
    \end{equation}
  \item harmonic oscillator potential \Ref{2.8b}:
    \begin{equation}
      \label{2.30}
    \begin{array}{rcl}
      \psi(x,t)&=&\ds(\cos 4at)^{-\frac{1}{2}} f\left(x^{-2}
        \cos4at\right)\\
      &&\ds\times \exp\left\{-i\Bigl(\frac{a}{2}x^2\tan 4at
        +\frac{\mu_3}{4}\ln\cos 4at -C\Bigr)\right\}\,;
    \end{array}
    \end{equation}
  \item anti-harmonic oscillator potential \Ref{2.8c}:
    \begin{equation}
      \label{2.31}
     \begin{array}{rcl}
       \psi(x,t)&=&\ds(\cosh 4at)^{-\frac{1}{2}} f\left(x^{-2}
         \cosh 4at\right)\\
       &&\ds\times \exp\left\{i\Bigl(\frac{a}{2}x^2\tan 4at-
         \frac{\mu_3}{4}\ln\cosh 4at + iC\Bigr)\right\}\,.
    \end{array}
    \end{equation}
  \end{enumerate}
Here $f$ is an arbitrary twice continuously differentiable function
and $C$ is an arbitrary constant.

\section{Conclusion}\label{5:sec}
As mentioned above {\sc Lie} symmetries of DG--equations considered are not
extensive enough to provide their linearizability by means of local
transformations. PDEs (\ref{e1}), (\ref{e2}) prove to be integrable
because of infinite {\em non-local\/} symmetries admitted. Take, as an
example, system (\ref{1.26}). It has been decoupled into a system of
two linear first-order PDEs (\ref{1.30}) by means of non-local
transformations of dependent and independent variables (\ref{1.27a}),
(\ref{1.29}), (\ref{1.29a}). It is well-known (see, e.g. \cite{Ovsian1:book})
that any linear first-order PDE admits an infinite parameter {\sc Lie}
transformation group. Consequently, system (\ref{1.30}) possesses an
infinite local symmetry. But after being rewritten in the initial
variables $\big(x,t,r(x,t),s(x,t)\big)$ it becomes non-local and can
not be found by using the infinitesimal {\sc Lie} algorithm.

In the case involved an existence of non-local symmetry was indicated
by a change of local symmetry of the DG--equation when the parameters
were specified to be (\ref{p1}), (\ref{p2}). These additional local
symmetries form the top of the \lq iceberg\rq , the main part of which
consists of non-local symmetries enabling us to integrate the
corresponding DG--equations.

Since we have the formulae for general solutions of systems of PDEs
(\ref{e1}), (\ref{e2}), it is not but natural to apply these to
analyze the initial value problem for these systems, which is
important for a physical interpretation of the equations. For example,
using formula (\ref{1.25}) it is not difficult to prove that the
initial value problem
\begin{eqnarray*}
  i\psi_t&=&\left\{(\mu_1-i) \Im\, {\psi_{xx}\over\psi}
          +2\left(\Im {\psi_x\over\psi}\right)^2 -
          \mu_1\Im\left({\psi_x\over\psi}\right)^2
          \right\}\psi,\\[1ex]
  \psi(x,0)&=&r_0(x)\exp\{is_0(x)\},
\end{eqnarray*}
where $r_0,s_0\in C^\infty(\rz,\rz)$ are arbitrary
functions such that $\mu_1s_0(x)\ne -\infty$, has a unique solution
given by the formula (\ref{1.25}), where $u(x,t)$ is a solution of the
initial value problem for the heat equation
\begin{displaymath}
u_t+\mu_1u_{xx}=0,\quad u(x,0)=\exp\left\{\frac{1}{\mu_1}s_0(x)\right\}
\end{displaymath}
and the function $f(y)$ reads
\begin{displaymath}
f(y)=R\Bigl(h(y)\Bigr)\exp\left\{-\frac{1}{\mu_1}S\Bigl( h(y) \Bigr)
\right\}.
\end{displaymath}
Here $h(y)$ is determined implicitly by the relation
\begin{displaymath}
y=\int_0^{h(y)}\exp\left\{\frac{2}{\mu_1}s_0(\tau)\right\}d\tau.
\end{displaymath}
But for the DG--equation (\ref{e1}) an analysis of the initial value
problem is complicated due to the complex structure of its general
solution.

The method of integration of DG--equations developed in the present paper
for the case of one space variable can be extended to a physically
more interesting case of three spatial dimensions. A principal idea of
such an extension is a utilization of generalized {\sc Euler-Amp\'ere}
transformations of the space $\big(t, \vec{x}, u(t,\vec{x}),
u_t(t,\vec x), \mbox{grad}\, u(t,\vec x)\big)$ suggested in
\cite{FuZhRe1}. The above transformations were used to study
compatibility and to construct a general solution of the
four-dimensional nonlinear {\sc d'Alembert}--eikonal system. This
problem is under investigation now and will be a topic of our future
publications.

\subsection*{Acknowledgments}
  We appreciate useful discussions with {\sc H.-D. Doebner}, {\sc
    W.I. Fushchych}, {\sc G.A.~Goldin}, and {\sc W.~L\"ucke} and we
  are grateful for their remarks and suggestions.
  One of us (RZ) would like to thank the ``Alexander von Humboldt
  Stiftung'' for financial support.

\end{document}